\documentclass[10 pt, twocolumn]{ieeeconf}  

\IEEEoverridecommandlockouts                              

\usepackage{hyperref,mathtools}

\mathtoolsset{showonlyrefs=true}
\usepackage{cite}
\let\labelindent\relax

\usepackage{amssymb,stmaryrd,graphicx,xcolor,enumitem}
\setlist[enumerate,1]{label={(\arabic*)}}
\DeclareMathOperator*{\argmax}{argmax}
\DeclareMathOperator*{\argmin}{argmin}

\DeclareMathOperator*{\subjectto}{\mathrm{subject~to}}
\DeclareMathOperator*{\sign}{\mathrm{sign}}
\DeclareMathOperator*{\G}{\mathbf{G}}
\DeclareMathOperator*{\F}{\mathbf{F}}

\newtheorem{theorem}{\protect\theoremname}
\newtheorem{corollary}{\protect\corollaryname}
\newtheorem{definition}{\protect\definitionname}
\newtheorem{remark}{\protect\remarkname}
\newtheorem{lemma}{\protect\lemmaname}

\newtheorem{assumption}{\protect\assumname}
\newtheorem{problem}{\protect\probname}

\providecommand{\definitionname}{\textbf{Definition}}

\providecommand{\remarkname}{\textbf{Remark}}
\providecommand{\theoremname}{\textbf{Theorem}}
\providecommand{\lemmaname}{\textbf{Lemma}}
\providecommand{\assumname}{\textbf{Assumption}}
\providecommand{\probname}{\textbf{Problem}}
\providecommand{\corollaryname}{\textbf{Corollary}}

\newcommand{\newsec}[1]{\vspace{0.2 cm} \noindent \textbf{#1}}

\title{\LARGE \bf
Formal Test Synthesis for Safety-Critical Autonomous Systems \\ based on Control Barrier Functions 
}

\author{Prithvi Akella, Mohamadreza Ahmadi, Richard M. Murray, and Aaron D. Ames$^{1}$
\thanks{$^*$ This work was supported by the Air Force Office of Scientific Research.}
\thanks{$^{1}$ The authors are with the California Institute of Technology, 1200 East California Boulevard, Pasadena, CA 91125, USA.
\href{mailto:pakella@caltech.edu}{\texttt{pakella@caltech.edu}},
\href{mailto:mrahmadi@caltech.edu}{\texttt{mrahmadi@caltech.edu}},
\href{mailto:murray@cds.caltech.edu}{\texttt{murray@cds.caltech.edu}},
\href{mailto:ames@caltech.edu}{\texttt{ames@caltech.edu}}}
}

\begin{document}

\maketitle

\begin{abstract}

The prolific rise in autonomous systems has led to questions regarding their safe instantiation in real-world scenarios.  Failures in safety-critical contexts such as human-robot interactions or even autonomous driving can ultimately lead to loss of life.  In this context, this paper aims to provide a method by which one can algorithmically test and evaluate an autonomous system.  Given a black-box autonomous system with some operational specifications, we construct a minimax problem based on control barrier functions to generate a family of test parameters designed to optimally evaluate whether the system can satisfy the specifications.  To illustrate our results, we utilize the Robotarium as a case study for an autonomous system that claims to satisfy waypoint navigation and obstacle avoidance simultaneously.  We demonstrate that the proposed test synthesis framework systematically finds those sequences of events (tests) that identify points of system failure.

\end{abstract}

\section{Introduction}
\nocite{video}
Autonomous systems have become increasingly pervasive in our everyday life, whether that be through the rise in interest for autonomous vehicles \cite{autonomous_vehicle}, intelligent defense systems \cite{Autonomous_Swarm_NAVY}, or even  human/robot interaction \cite{HRI_ahmadi}.  This rise in prevalence has motivated a similar increase in questions regarding the efficacy of these systems in safety critical contexts.  These questions are not entirely unfounded, however, as even in those cases when attempting to verify system efficacy, horrific accidents still occur \textit{e.g.} recent autonomous car crashes.  Nonetheless, the field is still pushing forward rapidly, and in the future, these autonomous systems will have to deal with even more complex, dynamic, and relatively unstructured environments.  Coupled with the cost of failure, this increase in system complexity makes systematic test and evaluation of these systems all the more necessary.


Significant work on this issue has been carried out by the test and evaluation (T\&E) community.   Reachability analysis has been used to shape critical test cases from existing data \cite{TE_model_based_unadaptive_1} .  At the discrete level, RRT has been used to efficiently search a feasible space to find critical sequences that identify failure of the underlying controller \cite{TE_model_based_unadaptive_2} .  Tests based on a graph-search framework over clustered, critical situations, have been developed via exhaustive mission simulation of the underlying system \cite{TE_model_based_unadaptive_3}.  Each of the aforementioned methods are model-based and not easily adaptable to other systems/testing environments as they are exhaustive.  To address the issue of adaptivity, one approach adaptively samples the feasible space to generate successively harder tests \cite{TE_specific_adaptive_1}.  That being said, the aforementioned frameworks require an accurate system model to function well, and except for the latter contribution, none are easily adaptable.  However, as noted in a memo by the Department of Defense \cite{DOD_article}, a testing framework that is both adaptive/adversarial and formally guarantees safety is still highly sought after.

Prior work in the T\&E community reference formal methods as a means by which one can formally guarantee safety/the lack thereof (see \cite{TE_formal_methods_CPS}).  Formal methods, specifically linear and signal temporal logic (LTL \& STL), have garnered significant interest in the controls community (see \cite{Safety_Logic1,Safety_Logic2,Safety_Logic3,Safety_Logic4,ahmadi2020barrier}).  In each of these cases, the logical specification encodes a control objective whose satisfaction is formally guaranteed via the barrier-based controller.  In this respect, control barrier functions are very useful in formally guaranteeing these logical specifications insofar as satisfying a specification and remaining within a safe set are both set-based arguments \cite{TAC_Paper}.  However, these formal guarantees require specific knowledge of the onboard controller and system dynamics - for the test engineer, this is oftentimes not the case.

\begin{figure}
    \centering
    \includegraphics[width=0.49\textwidth]{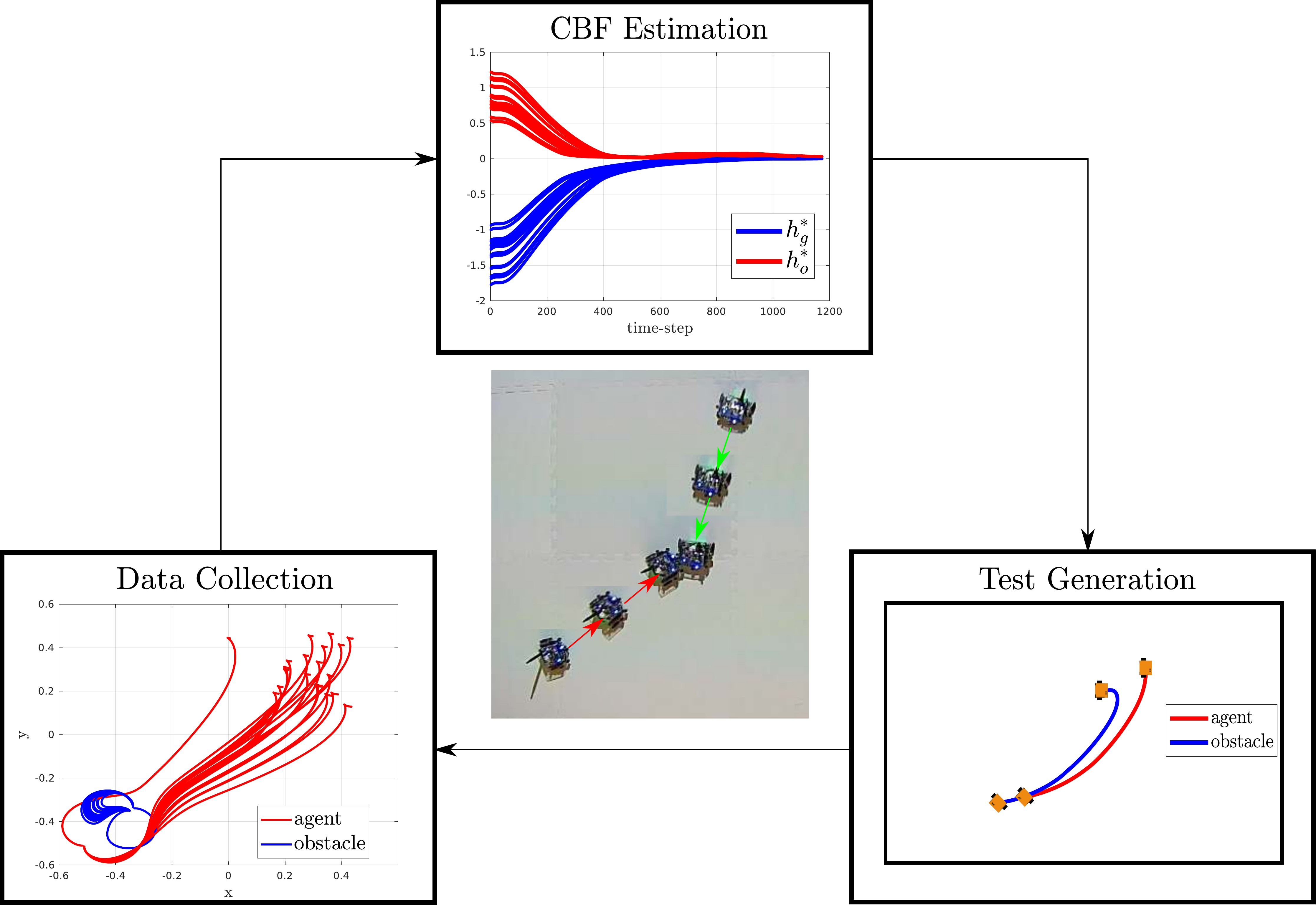}
    \caption{The flowchart for the proposed test generation framework.  The framework starts at (bottom left) collecting data for the system to be tested in order to (center top) estimate CBFs corresponding to the control objectives the system intends to satisfy.  These estimated CBFs are used in a minimax game to (bottom right) generate tests designed to verify system efficacy in satisfying the aforementioned objective.  This test generation framework is designed to systematically identify (center middle) points of system failure that may occur during general operation.}
    \label{fig::Title_flowchart}
\end{figure}

\newsec{Our Contribution:} In this paper, the overarching goal is to start to bridge the work done in the controls and the T\&E community.  Specifically, we address the issue of designing an adaptable/adversarial testing framework. Given an autonomous system with some operational specifications, we construct a minimax problem whose solution defines testing scenarios intended to optimally frustrate satisfaction of the given specifications without specific knowledge of the onboard control architecture.  To this end, we begin by collecting data of the autonomous system satisfying the specifications.  Then, we use the collected demonstration data to frame Linear Programs that develop approximate Control Barrier Functions corresponding to the operational specifications of the autonomous system. Finally, we use these approximate control barrier functions to develop a minimax game to solve for optimal testing parameters designed to frustrate satisfaction of the specifications.  The proposed method is illustrated in Figure~\ref{fig::Title_flowchart}.

\newsec{Outline:} In Section~\ref{sec:probform}, we review some preliminary definitions and  formally define the problem under study.  In Section~\ref{sec::Main_Result}, we detail the main result of the paper, \textit{i.e.,} a minimax game for test generation. In Section~\ref{sec::corollaries}, we couple the result with a linear program to systematically generate difficult tests. Finally, in Section~\ref{sec::simulations_and_experiments}, we illustrate our proposed methodology with a case study.

\section{Problem Formulation} \label{sec:probform}

In this section, we present some  notions used in the sequel and formally define the problem under study.

\subsection{Preliminaries}
\label{sec::problem_setup}
We consider a class of systems (to-be-tested) that can be modeled as a dynamical system with affine inputs:
\begin{equation}
    \label{dyn_sys}
    \dot{x} = f(x) + g(x)u, \quad x \in \mathcal{X} \subset \mathbb{R}^n, \quad u \in \mathcal{U} \subset \mathbb{R}^m.
\end{equation}
Furthermore, we will assume that both $f(x)$ and $g(x)$ are locally Lipschitz.  For any function $h(x)$,
\begin{align}
    L_fh(x) &\triangleq \nabla_xh(x)f(x), \\
    L_gh(x) &\triangleq \nabla_xh(x)g(x),
\end{align}
are its Lie derivatives.  

\newsec{Formal Methods:} We will define $\mathcal{A}$ to be the set of atomic propositions from which the provided control objective, \textit{i.e.,} a temporal logic specification, has been synthesized.  We use the following notation to represent the truth/lack thereof for an atomic proposition
\begin{equation}
    \forall \phi \in \mathcal{A}, \quad \llbracket \phi \rrbracket \triangleq \{x \in \mathcal{X} | \phi(x) = \mathrm{TRUE} \},
\end{equation}
where $\phi(x)$ denotes the atomic proposition evaluated at the state $x$.  In addition, we will define the symbols $\neg, \wedge, \lor$ to correspond to  negation, conjunction, and disjunction respectively.  That is, $\neg \phi = $ TRUE when $\phi = $ FALSE.  Likewise $\phi \wedge \omega =$ TRUE when $\phi = $ TRUE and $\omega = $ TRUE, and $\phi \lor \omega =$ TRUE when either $\phi=$ TRUE or $\omega = $ TRUE.

In this paper, we consider a subset of temporal logic (TL) operators, $\F$uture and $\G$lobal, defined as follows (here $\equiv$ denotes a logical equivalency):
\begin{align}
    \F\phi & \equiv \exists~t^*\geq 0~\mathrm{s.t.~}x(t^*)\in\llbracket \phi \rrbracket,\\
    \G\phi & \equiv \forall~t\geq 0,~x(t)\in\llbracket \phi \rrbracket.
\end{align}
While this seems restrictive, these two operators can be composed to consider more complex LTL specifications, such as $\square \lozenge \phi \equiv \G(\F\phi)$.  

\newsec{Control Barrier Functions (CBF):} To provide a metric by which we measure satisfaction of the provided specification, we will establish a correspondence between these TL specifications and control barrier functions, $h$.  To start, we first define extended class-$\mathcal{K}$ functions, $\alpha: (-b,a) \to (-\infty,\infty)$, to be those functions, $\alpha$, that are strictly increasing and satisfy $\alpha(0) = 0$.  Here, $a,b>0$.  Using these extended class-$\mathcal{K}$ functions, we can define Control Barrier Functions (CBF).
\vspace{.01cm}
\begin{definition}
[Control Barrier Functions (CBF)]
\label{def::cbf}
\textit{For a dynamical system of the form \eqref{dyn_sys}, a differentiable function, $h: \mathbb{R}^n \to \mathbb{R}$ is considered a control barrier function if it satisfies the following criteria:
\begin{equation}
    \label{}
    \sup\limits_{u\in\mathcal{U}} \left[L_fh(x) + L_gh(x)u + \alpha(h(x)) \right] \geq 0,\quad \forall x \in \mathcal{X},
\end{equation}
where $\alpha$ is an extended class-$\mathcal{K}$ function~\cite{TAC_Paper}.}
\end{definition}
\vspace{.2cm}
The usefulness of a CBF is in guaranteeing the forward invariance of its 0-superlevel set:
\begin{align}
    & \mathcal{C}_h & & \hspace{-0.9 in}= \{x \in \mathbb{R}^n~|~ h(x) \geq 0 \}, \\
    & \partial\mathcal{C}_h & &\hspace{-0.9 in}= \{x \in \mathbb{R}^n~|~ h(x) = 0 \}.
\end{align}
Indeed, it was shown in Proposition 1 of \cite{TAC_Paper} that a CBF, as in Definition~\ref{def::cbf}, guarantees forward invariance of its 0-Superlevel set, $\mathcal{C}_h$.  Here, we note that what we call control barrier functions are termed as \textit{zeroing control barrier functions} in \cite{wang2017safety}.  Finally, a finite time convergence control barrier function requires $\alpha(x) =\gamma\sign(x)|x|^\rho$ to ensure finite time convergence to the set, $\mathcal{C}_h$, by $T = \frac{1}{\gamma(1-\rho)}|h(x_0)|^{1-\rho}$, provided $h(x_0) \leq 0$ \cite{Finite_CBF}.

\subsection{Problem Statement}
\label{sec::problem_statement}
As mentioned earlier, the overarching test and evaluation goal is to validate an autonomous system's capacity to satisfy a provided TL specification.  However, as we have no knowledge of the controller on-board the system to-be-tested, not only do we have no metric of quantifying success for the TL specification, but we also do not have a systematic method of developing difficult tests by which to identify control system failures in satisfying the specification.  We will show in the sequel that there exists a correspondence between CBFs and TL specifications. So, if we could determine these CBFs for the system at hand, we can use them to test the system against a given specification.  This chain of reasoning is the basis for Figure~\ref{fig::Title_flowchart}.  To that effect, we collect the following experimental data of the system satisfying the  control objective:
\vspace{0.2cm}
\begin{definition}
    [Data-Set]
    \label{demonstrations}
\textit{    Define $\mathbb{D}_i = \{ (x^i_k,u^i_k) \in \mathbb{R}^n\times\mathbb{R}^m~|~k = 0,1,\dots,T_i \}$ as the data-set of state, action pairs for demonstration, $i$.  Here, $k$ indexes time until $T_i$, which is the max time for the specific demonstration at hand.  Then, define $\mathbb{D} = \{ \mathbb{D}_1, \dots, \mathbb{D}_r \}$ as the set of all provided demonstrations.}
\end{definition}
\vspace{0.2cm}
\begin{assumption}
    \label{Labeling}
    For the provided data-set, $\mathbb{D}$, and associated specification, the data-set for each demonstration, $\mathbb{D}_i$, terminates when the specification is satisfied.  \textit{e.g.} for a specification defined as $\F\phi \wedge \G\omega$, where $\phi,\omega \in \mathcal{A}$, then for each $\mathbb{D}_i$,
    \begin{itemize}
    \item $x^i_{T_i} \in \llbracket \phi \rrbracket$ and $x^i_k \not \in \llbracket \phi \rrbracket$ for all $k = 0,1,\dots,T_i-1$, and
    \item $x^i_k \in \llbracket \omega \rrbracket$ for all $k = 0,1,\dots,T_i$.
\end{itemize}
\end{assumption}
\vspace{0.2cm}
We  use the generated data-set, $\mathbb{D}$, to determine composite CBFs that mimic system behavior.  We compose these CBFs from a candidate set of barrier functions defined as follows:
\vspace{.0cm}
\begin{definition}
    [Candidate Barrier Set] \textit{We call
    \begin{equation*}
        \mathcal{B} \triangleq \{h_1, h_2, \dots, h_q \},
    \end{equation*}
     a candidate barrier set for some provided, continuously differentiable functions, $\{h_i\}_{i=1}^q$.} 
\end{definition}
\vspace{.2cm}
Note that in the above definition, each component of the candidate barrier set may not be a valid CBF, \textit{i.e.} $\mathcal{B}$ could be the set of all polynomials of degree, $n\leq q-1$.  Finally, we need to formalize how we specifically identify these testing scenarios. \vspace{.2cm}
\begin{definition}
[Testing Parameters]
\label{testing_parameters}
\textit{We define the vector, $d\in\mathbb{R}^p$, to be a collection of testing parameters used to generate tests \textit{e.g.} the location of obstacles, time when a phenomena starts, \textit{etc}.} 
\end{definition}
\vspace{.2cm}
With these definitions in place, the problem statement is as follows: 
\vspace{.2cm}
\begin{problem}
\label{main_problem}
\textit{Given an autonomous system whose controller is unknown, $\mathbb{D}$, $\mathcal{B}$, and a TL specification the system intends to satisfy, devise an adaptive/adversarial strategy to identify a set of testing parameters $d$. }
\end{problem}
\vspace{0.2cm}
We show in the next section that these test parameters $d$ characterize a test scenario designed to validate that the autonomous system reliably satisfies a given TL specification.


\section{Main Result}
\label{sec::Main_Result}
This section will detail the main result of this paper - the minimax game formulated to generate optimal test parameters, $d^*$, designed to frustrate satisfaction of a TL specification expressed through CBFs.
\subsection{Main Result}
To preface the main result, we will make the following remark to simplify notation:
\vspace{.2cm}
\begin{remark}
\label{approximate_barrier_labeling}\textit{We denote $h^F_i~,i\in \mathcal{I}$ to be a set of  CBFs for a finite number of specifications of the type $\mathbf{F}\phi_i$.  Likewise, $h^G_j,~j \in \mathcal{J}$ denote CBFs for specifications of the type $\mathbf{G}\omega_j$.  That is, $\mathcal{C}_{h^F_i} \equiv \llbracket \phi_i \rrbracket$, $\forall~ i\in\mathcal{I}$, and $\mathcal{C}_{h^G_j} \equiv \llbracket \omega_j \rrbracket$, $\forall~j\in\mathcal{J}$.}
\end{remark}
\vspace{.2cm}

In addition, we will make the following assumption to simplify the formulations in the sequel.
\vspace{.2cm}
\begin{assumption}
\label{test_restriction}
\textit{We will assume that the CBFs $h^G_j,~j\in \mathcal{J}$ depend on a set of test parameters $d$.  That is, $h^G_j:\mathbb{R}^n\times\mathbb{R}^p \to \mathbb{R}$ and $\dot{h}^G_j: \mathbb{R}^n\times\mathbb{R}^p\times\mathbb{R}^m \to\mathbb{R}$ whereas, $h^F_i: \mathbb{R}^n \to \mathbb{R}$.}
\end{assumption}
\vspace{.2cm}

We will define the following set of feasible inputs:
\begin{align}
    & \mathcal{U}(x,d) = \label{eqn::feasible_set} \\
    & \{ u \in \mathcal{U}~|~\dot{h}^G_j(x,u,d) \geq -\alpha_j( h^G_j(x,d)),~\forall~j\in\mathcal{J} \}, \nonumber 
\end{align}
where each $\alpha_j$ is the corresponding extended class-$\mathcal{K}$ function with respect to which $h^G_j$ is a CBF.  Likewise, we will define:
\begin{equation}
    x(t)|_{u(t)} \triangleq x(0) + \int_0^t \left(f(x(s))+g(x(s))u(s)\right) ds,
\end{equation}
to be the solution to equation~\eqref{dyn_sys} provided the input signal, $u(t)$.

Likewise, we will make the following assumption to frame the type of specifications accounted for by the testing framework to be detailed:
\vspace{0.2 cm}
\begin{assumption}
We assume that the provided TL specification can be recast into the following form:
\begin{equation}
    \label{eqn::sys_specification}
    \left[\lor_{i\in\mathcal{I}} \left( \F \phi_i \right) \right] \wedge \left[\wedge_{j\in\mathcal{J}} \left( \G\omega_j \right) \right],~\phi_i,\omega_j\in\mathcal{A}~\forall~i,j,
\end{equation}
with the following initial conditions:
\begin{subequations}
\begin{align}
    & \lor_{i\in\mathcal{I}}(\phi_i(x(0))) & & \hspace{-0.8 in} =\mathrm{FALSE}, \label{eqn::initially_not_F}\\
    & \wedge_{j\in\mathcal{J}}(\omega_j(x(0))) & & \hspace{-0.8 in} =\mathrm{TRUE}. \label{eqn::initially_G}
\end{align}
\end{subequations}
\end{assumption}
\vspace{0.2cm}
Intuitively, specifications of type~\eqref{eqn::sys_specification} denote control objectives wherein the system must ensure continued satisfaction of multiple control objectives while accomplishing at least one of a subset of tasks \textit{e.g.} navigating to one of multiple waypoints while avoiding all obstacles. Equations~\eqref{eqn::initially_not_F} and \eqref{eqn::initially_G} indicate that the system does not start in trivial states, wherein the specification~\eqref{eqn::sys_specification} has already been satisfied.  Finally, to account for an adversarial testing framework, we specify that the test parameters are a function of the current state, \textit{i.e.} $d(x)$, where the specific functional form is expressed in Theorem~\ref{algorithmic_test_generation}.  Intuitively, the idea is that for tests to be adversarial to system action, they must, necessarily, depend on the system state.

Under the notation specified in Remark~\ref{approximate_barrier_labeling}, the main result is as follows:
\vspace{.2cm}
\begin{theorem}
[Algorithmic Test Generation]
\label{algorithmic_test_generation}
\textit{Given an autonomous system and a TL specification of the form in equation~\eqref{eqn::sys_specification}, 
the solution, $d^*(x)$, to the minimax game:
\begin{align}
    d^*(x) = & \,\,\,\, \argmin\limits_{d \in \mathbb{R}^p} & & \hspace{-0.3 in}\max\limits_{u \in \mathcal{U}(x,d)} \, \, \sum\limits_{i \in \mathcal{I}} \dot{h}^F_i(x,u), 
    \label{differential_game}\quad \quad \quad  \tag{Minimax}
\end{align}
defines an optimal test parameter sequence, $d^*(x(t))$, predicated on a state trajectory, $x(t)|_{u(t)}$, and the control signal, $u(t)$, \textit{i.e.,}  $d^*(x(t))$ identifies a sequence of test scenarios designed to ensure system satisfaction of the following specification:}
\begin{equation}
    \label{eqn::d_specification}
    \left[ \wedge_{i\in\mathcal{I}} \left(\G \neg \phi_i \right) \right] \lor \left[ \lor_{j\in\mathcal{J}}
    \left( \F \neg \omega_j\right)\right],~\phi_i,\omega_j\in\mathcal{A}~\forall~i,j. 
\end{equation}
\end{theorem}

\subsection{Proof of Main Result}
This section contains the lemmas necessary to prove the main result, Theorem~\ref{algorithmic_test_generation}.  For all maximization/minimization problems contained within, we specify that infeasibility of the associated optimization problem corresponds to a value of $-\infty,\infty$ respectively.

To start, we need to show that TL specification~\eqref{eqn::d_specification} and TL specification~\eqref{eqn::sys_specification} are mutually exclusive.  To that end, we have the following Lemma regarding relations between TL operators:
\vspace{.2cm}
\begin{lemma}
    \label{lemma::TL_operator_relations}
\textit{    The following relations are true:
    \begin{align}
        \neg \G \phi & \equiv \F (\neg \phi), \label{eqn::TL_relation_2}\\
        \neg \F \phi & \equiv \G (\neg \phi). \label{eqn::TL_relation_3}
    \end{align}
    }
\end{lemma}
\vspace{.2cm}
\begin{proof}
For equation~\eqref{eqn::TL_relation_2},
\begin{equation*}
    \neg \G \phi \equiv \exists t^*\geq 0~|~x(t^*)\in\llbracket \neg \phi \rrbracket \equiv \F(\neg\phi).
\end{equation*}
Likewise, for equation~\eqref{eqn::TL_relation_3},
\begin{equation*}
    \neg\F\phi \equiv \forall~t\geq0~x(t)\in\llbracket\neg\phi\rrbracket \equiv \G(\neg \phi).
\end{equation*}
\end{proof}

Using Lemma~\ref{lemma::TL_operator_relations} and De Morgan's Law, we can prove that the two TL specifications, \eqref{eqn::d_specification} and \eqref{eqn::sys_specification}, are mutually exclusive:
\vspace{.1cm}
\begin{lemma}
    \label{lemma::mutual_exclusivity}
\textit{    TL specifications \eqref{eqn::d_specification} and \eqref{eqn::sys_specification} are mutually exclusive.}
\end{lemma}
\vspace{.2cm}
\begin{proof}
    \begin{align*}
        & \neg \left[ \left[\lor_i \left( \F \phi_i \right) \right] \wedge \left[\wedge_j \left( \G\omega_j \right) \right] \right] \\
        \equiv & \neg \left[\lor_i \left( \F \phi_i \right) \right] \lor \neg \left[\wedge_j \left( \G\omega_j \right) \right] \\
        \equiv & \left[\wedge_i \neg(\F\phi_i) \right] \lor \left[ \lor_j \neg(\G\omega_j)\right] \\
        \equiv & \left[ \wedge_i (\G\neg\phi_i)\right] \lor \left[ \lor_j (\F\neg \omega_j)\right]
    \end{align*}
\end{proof}

Effectively, Lemma~\ref{lemma::mutual_exclusivity} proves that if $d^*(x(t))$ ensures system satisfaction of TL specification~\eqref{eqn::d_specification}, then the sequence of test parameters did identify a system failure insofar as the system failed to satisfy the  specification~\eqref{eqn::sys_specification}.  It remains, however, to show that minimax game~\eqref{differential_game} defines a sequence, $d^*(x(t))$, that forces the system to satisfy~\eqref{eqn::d_specification}.  To that end, we have the following Lemma that draws a correspondence between CBFs and TL specifications:
\vspace{.2cm}
\begin{lemma}
    \label{equivalence_logic_safety}
  \textit{  For an atomic proposition, $\phi \in \mathcal{A}$, if there exists a function, $h_\phi(x)$, such that $\mathcal{C}_{h_\phi} = \llbracket \phi \rrbracket$, then:
    \begin{equation*}
        \G\phi \equiv h_\phi(x(t)) \geq 0, ~\forall~ t \geq 0,
    \end{equation*}
    and:
    \begin{equation*}
        \F\phi \equiv \exists~t^* <\infty~\mathrm{s.t.}~h_\phi(x(t^*)) \geq 0.
    \end{equation*}
    Furthermore, if $h_\phi(x)$ is a CBF, then $\exists~u(t)$ such that $\G\phi = $TRUE.   Likewise, if $h_\phi(x)$ is an FTCBF, then $\exists~u(t)$ such that $\F\phi = $TRUE.}
\end{lemma}
\vspace{.2cm}
\begin{proof}
    For $\F\phi$:
    \begin{align*}
        \F\phi & \quad \equiv \quad \exists~0\leq t^* <\infty~\mathrm{s.t.}~x(t^*) \in \llbracket \phi \rrbracket, \nonumber \\
        & \quad \equiv \quad \exists~0\leq t^* <\infty~\mathrm{s.t.}~ x(t^*) \in \mathcal{C}_{h_\phi}, \nonumber \\
        & \quad \equiv \quad \exists~0\leq t^* <\infty~\mathrm{s.t.}~ h_\phi(x(t^*)) \geq 0.
    \end{align*}
    Hence, if $h_\phi(x)$ is an FTCBF wherein $h(x(0)) \leq 0$, then an input sequence, $u(t)$, that satisfies:
    \begin{align*}
        & L_fh(x(t)) + L_gh(x(t))u(t) + \gamma \sign(h(x(t)))\left| h(x(t))\right|^\rho \geq 0, \\
        & \quad \quad \quad \forall~ t \leq T= \frac{1}{\gamma(1-\rho)}|h(x_0)|^{1-\rho}
    \end{align*}
    ensures $
       h(x(T))\geq 0 \implies \F\phi = \mathrm{TRUE}.
    $ 
    $\G\phi$ follows similarly.
\end{proof}

Lemma~\ref{equivalence_logic_safety} provides a metric by which to verify that $d^*(x(t))$ ensures system satisfaction of  specification~\eqref{eqn::d_specification}.  Specifically, Lemma~\ref{equivalence_logic_safety} requires that $d^*(x(t))$ either ensure $h^F_i(x(t)) < 0$ $\forall~i \in \mathcal{I}$ and $\forall~t\geq0$, or $h^G_j(x(t)) < 0$ for at least one $j\in\mathcal{J}$ and $t\geq 0$.  To show this, we require the following definitions for the optimal cost, $s$, optimal input, $u^*$, and optimal test parameter, $d^*$:
\begin{align}
    & \,\,\, s(x(t),d) & & \hspace{-0.15 in} = & & \hspace{-0.1 in}\max\limits_{u\in\mathcal{U}(x(t),d)}  & & \hspace{-0.1 in} \sum\limits_{i\in\mathcal{I}}\dot{h}^F_i(x(t),u), \label{eqn::max_derivative}\\
    & u^*(x(t),d) & & \hspace{-0.15 in} = & & \hspace{-0.1 in}\argmax\limits_{u\in\mathcal{U}(x(t),d)}  & & \hspace{-0.1 in} \sum\limits_{i\in\mathcal{I}}\dot{h}^F_i(x(t),u), && \label{eqn::u_optimal}\\
    & \,\,\,\,d^*(x(t)) & & \hspace{-0.15 in} = & & \hspace{-0.1 in}\,\,\,\,\argmin\limits_{d\in\mathbb{R}^p}  & & \hspace{-0.1 in} \sum\limits_{i\in\mathcal{I}}\dot{h}^F_i(x(t),u^*(x(t),d)).  && \label{eqn::d_optimal}
\end{align}
Here, we note that equation~\eqref{eqn::d_optimal} is a re-casting of equation~\eqref{differential_game} accounting for the optimal input, $u^*(x(t),d)$.  In addition, we will define the following set of invalidating test parameters:
\begin{equation}
    \label{eqn::infeasibility_set}
    \mathcal{D}(x) = \{d \in \mathbb{R}^p~|~\mathcal{U}(x,d)=\varnothing\}.
\end{equation}
With the above definitions, we have the following Lemma:
\vspace{.2cm}
\begin{lemma}
    \label{lemma::ensuring_infeasibility}
\textit{If, for some $x(t)$, $\mathcal{D}(x(t)) \neq \varnothing$, then the optimal solution, $d^*$, to equation~\eqref{differential_game} is such that, $d^*\in\mathcal{D}(x(t))$.}
\end{lemma}
\vspace{.2cm}
\begin{proof}
    First, we note that,
    \begin{equation}
        \label{eqn::infeasibility_yields_infinity}
        \forall~d \in \mathcal{D}(x(t)),~s(x(t),d) = -\infty.
    \end{equation}
    The equation above comes from the infeasibility of maximization problem~\eqref{eqn::max_derivative}, which results in a value of $s = -\infty$. Furthermore, equation~\eqref{differential_game} is equivalent to:
    \begin{align}
        \label{eqn::recasting_dstar_again}
        d^*(x(t)) = & \,\,\,\, \argmin\limits_{d \in \mathbb{R}^p} & & \hspace{-0.6 in} s(x(t),d).
    \end{align}
    Based on the Locally Lipschitz assumptions made for $f(x)$ and $g(x)$ in equation~\eqref{dyn_sys} and the requirement that a CBF, $h(x)$, is differentiable at least once, it is true that
    \begin{equation*}
        L_fh^F_i(x),L_gh^F_i(x)~\mathrm{are~bounded}~\forall~i\in\mathcal{I}.
    \end{equation*}
    In addition,
    \begin{equation*}
        \forall~u\in\mathcal{U}(x(t),d),~u~\mathrm{is~bounded}.
    \end{equation*}
    Therefore,
    \begin{equation*}
        \dot{h}^F_i(x(t),u) = L_fh(x(t)) + L_gh(x(t))u~\mathrm{is~bounded}~\forall~i\in\mathcal{I}.
    \end{equation*}
    As defined in equations~\eqref{eqn::max_derivative} and \eqref{eqn::u_optimal}, it is also true that
    \begin{equation*}
        s(x(t),d) = \sum\limits_{i\in\mathcal{I}}\dot{h}^F_i(x(t),u^*(x(t),d)).
    \end{equation*}
    As each component, $\dot{h}^F_i$, is finite and $|\mathcal{I}|<\infty$, the following is true:
    \begin{equation}
        \label{eqn::feasibility_yields_finite}
        \exists~M < \infty ~\mathrm{s.t.}~\left|s(x(t),d)\right|<M, \quad \forall~d\not\in\mathcal{D}(x(t)).
    \end{equation}
    By definition of $\argmin$ and using equations~\eqref{eqn::infeasibility_yields_infinity}, \eqref{eqn::recasting_dstar_again}, and~\eqref{eqn::feasibility_yields_finite}, we have that
 $
        d^*(x(t))\in\mathcal{D}(x(t)).$
\end{proof}

With Lemma~\ref{lemma::ensuring_infeasibility}, we can show that the sequence, $d^*(x(t))$, attempts to force the system to satisfy, $\lor_j (\F\neg \omega_j)$.  We will show this first for a single $\G\omega$:
\vspace{.2cm}
\begin{lemma}
    \label{lemma::invalidity_Gomega}
\textit{    If, for a given state trajectory, $x(t)$, $\omega(x(0)) = $ TRUE, and $\mathcal{D}(x(t)) \neq \varnothing$ $\forall$ $t\geq0$ with $|\mathcal{J}| = 1$, then:
    \begin{equation}
        \label{eqn::finite_time_invalidation_omega}
            \forall~\delta>0,~\exists~t^*_\delta \in (0,\infty)~\mathrm{s.t.}~h_\omega(x(t^*),d^*d(t^*)) < \delta,
    \end{equation}
    where $h_\omega$ is the CBF corresponding to $\G\omega$.}
\end{lemma}
\vspace{.2cm}
\begin{proof}
    Via Lemma~\ref{lemma::ensuring_infeasibility}, we know that 
$
        \forall~t\geq 0,~d^*(x(t)) \in \mathcal{D}(x(t)).
$ 
    As $|\mathcal{J}|=1$, this implies that $\forall~t\geq0$,
    \begin{equation*}
        \label{eqn::no_u_exists}
        \dot{h}_\omega(x(t),u,d^*(x(t))) < -\alpha(h_\omega(x(t),d^*(x(t)))),~\forall~u\in\mathcal{U}.
    \end{equation*}
    As $\alpha(\cdot)\in\mathcal{K}$ (abbreviating $d^*(x(t))$ to $d^*(t)$):
    \begin{equation*}
        h_\omega(x(t),d^*(t)) < \beta(h_\omega(x(0),d^*(0)),t),
    \end{equation*}
    where $\beta(\cdot)$ is a class-$\mathcal{KL}$ function.  As a result:
    \begin{equation*}
        \exists~t^*\in(0,\infty)~\mathrm{s.t.}~\beta(h_\omega(x(0),d^*(0)),t^*) \leq \delta,
    \end{equation*}
    choosing $t_\delta^* = t^*$ completes the proof. 
\end{proof}

Lemma~\ref{lemma::invalidity_Gomega} is why we specify that the sequence, $d^*(x(t))$, attempts to force system satisfaction of $\lor_j(\F\neg\omega_j)$ as opposed to specifying that it guarantees that the system will satisfy the same specification. As minimax game~\eqref{differential_game} constrains system action, $u$, to ensure $\wedge_j(\G\omega_j)$, the test sequence can only get $\delta$ close to invalidation assuming optimal system action. This discrepancy will be made clear when compared with Lemma~\ref{lemma::invalidation_Fphi}:
\vspace{.2cm}
\begin{lemma}
    \label{lemma::invalidation_Fphi}
\textit{    If $\phi(x(0)) = $ FALSE and $|\mathcal{I}|=1$, then the test parameter sequence, $d^*(x(t))$, is guaranteed to find a system trajectory, $x(t)|_{u^*(x(t),d^*(x(t)))}$, that satisfies $\G\neg\phi$ provided a trajectory exists wherein:
    \begin{align}
        \dot{h}_\phi(x(t),u^*(x(t),d(t))) & \leq 0,~\forall~t\geq 0, \label{eqn::h_phi_always_decreasing}\\
        \mathcal{D}(x(t)) & = \varnothing,~\forall~t\geq0, \label{eqn::no_invalidity_set}
    \end{align}
    for some $d(t)$.}
\end{lemma}
\vspace{.2cm}
\begin{proof}
    First, we denote $h_\phi(x)$ to be the CBF corresponding to $\F \phi$.  It follows from Lemma~\ref{equivalence_logic_safety} then,
    \begin{equation}
        \label{eqn::phi_false_iff_negative}
        \phi(x(0)) = \mathrm{FALSE} \equiv h_\phi(x(0)) < 0.
    \end{equation}
    From equation~\eqref{eqn::phi_false_iff_negative}, to prove $\G\neg\phi$, it is sufficient to prove:
    \begin{equation}
        \label{eqn::general_h_always_decreasing}
        \dot{h}_\phi(x(t),u(t)) \leq 0, \quad \forall~t\geq 0,
    \end{equation}
    as if true:
    \begin{align*}
        h_\phi(x(t)) & = h_\phi(x(0)) + \int_0^t \dot{h}_\phi(x(s),u(s)) \mathrm{ds}, \\
        & < \int_0^t \dot{h}_\phi(x(s),u(s)) \mathrm{ds} \leq 0, \\
        & \implies x(t)|_{u(t)} \in \llbracket \neg \phi \rrbracket,~\forall~t\geq 0~ \equiv \G\neg\phi.
    \end{align*}
    As a result, all that remains is to show that equation~\eqref{eqn::h_phi_always_decreasing} is satisfied by $d^*(x(t))$.  Here, equation~\eqref{eqn::no_invalidity_set} ensures that the results of Lemma~\ref{lemma::ensuring_infeasibility} do not apply, as otherwise $d^*(x(t))\in\mathcal{D}(x(t))$ and we cannot make a statement regarding $s(x(t),d^*)$.  Then, by definition of $\argmin$ and equation~\eqref{eqn::d_optimal} (abbreviating $d^*(x(t))$ to $d^*(t)$):
    \begin{equation*}
        \dot{h}_\phi(x(t),u^*(x(t),d^*(t))) \leq \dot{h}_\phi(x(t),u^*(x(t),d(t))),
    \end{equation*}
    which results in:
    \begin{equation}
        \label{eqn::h_decreasing_under_optimality}
        \dot{h}_\phi(x(t),u^*(x(t),d^*(t))) \leq 0.
    \end{equation}
    From equation~\eqref{eqn::h_decreasing_under_optimality} and the sufficiency proof predicated on  equation~\eqref{eqn::general_h_always_decreasing}, we have:
    \begin{equation*}
        x(t)|_{u^*(x(t),d^*(t))} \in \llbracket \neg \phi \rrbracket,~\forall~t\geq 0 \equiv \G\neg\phi.
    \end{equation*}
\end{proof}

With the aforementioned lemmas, we are now ready to prove Theorem~\ref{algorithmic_test_generation}. 

\begin{proof}
[Theorem~\ref{algorithmic_test_generation}] If both $|\mathcal{I}| = 1$ and $|\mathcal{J}|=1$, then the result stems directly from Lemmas~\ref{lemma::invalidity_Gomega} and \ref{lemma::invalidation_Fphi}.  First we note the following is true:
\begin{equation}
    \label{eqn::vacuously_true_D}
   \left( \mathcal{D}(x(t)) = \varnothing \right) \lor \left( \mathcal{D}(x(t)) \neq \varnothing \right) = \mathrm{TRUE},~\forall~t\geq 0.
\end{equation}
As a result, it is true that $\forall~t \geq 0$, the optimal test parameter sequence, $d^*(x(t))$, attempts to ensure that the following is true:
\begin{align}
    & \left(\dot{h}_\phi(x(t),u^*(x(t),d^*(x(t))))\leq 0 \right) \lor \label{eqn::CBF_inequality_forall_time} \\
    & \left(\dot{h}_\omega(x(t),u,d^*(x(t))) <  -\alpha(h_\omega(x(t),d^*(x(t)))~\forall~u\in\mathcal{U}\right)  \nonumber \\
    & \quad \quad \quad \forall~t\geq 0.\nonumber
\end{align}
Hence, if either $\mathcal{D}(x(t))=\varnothing$ or $\mathcal{D}(x(t)) \neq \varnothing$ persist $\forall~t\geq 0$, then the the results of Lemmas~\ref{lemma::invalidity_Gomega} and \ref{lemma::invalidation_Fphi} ensure $d^*(x(t))$ attempts to force the system to satisfy, $
    \G \neg \phi \lor \F \neg \omega.$ 
We lose the guarantee on $\G\neg\phi$ that we had in Lemma~\ref{lemma::invalidation_Fphi} as we can no longer ensure $\mathcal{D}(x(t)) = \varnothing$, $\forall~t\geq0.$  However, whenever $\mathcal{D}(x(t)) = \varnothing$, $d^*(x(t))$ will steer the system away from achieving $\F\phi$, if feasible.  

This same rationale extends to the case wherein $|\mathcal{I}| \neq 1$ and/or $|\mathcal{J}|\neq1$.  Since~\eqref{eqn::vacuously_true_D} holds, for the multi-specification case, $d^*(x(t))$ attempts to ensure:
\begin{align}
    & \left(\sum_{i\in\mathcal{I}}\dot{h}^F_i(x(t),u^*(x(t),d^*(x(t))))\leq 0 \right) \lor \label{eqn::multi_CBF_inequality_forall_time} \\
    & \left(\dot{h}^G_j(x(t),u,d^*(x(t))) <  -\alpha_j(h^G_j(x(t),d^*(x(t)))~\forall~u\in\mathcal{U}\right)  \nonumber \\
    & \quad \quad \quad \forall~t\geq 0,~\mathrm{and~for~at~least~one~}j, \nonumber
\end{align}
For the first inequality in equation~\eqref{eqn::multi_CBF_inequality_forall_time}, the following implication is true:
\begin{align}
    & \sum_{i\in\mathcal{I}}\dot{h}^F_i(x(t),u^*(x(t),d^*(x(t))))\leq 0 \implies \label{eqn::decrease_in_at_least_one_phi}\\
    & \lor_{i\in\mathcal{I}}\left( \dot{h}^F_i(x(t),u^*(x(t),d^*(x(t)))) \leq 0\right) = \mathrm{TRUE}. \nonumber 
\end{align}
Implication~\eqref{eqn::decrease_in_at_least_one_phi} can be deduced from a contradiction.  If, for the same implication, we were to assume the LHS of~\eqref{eqn::decrease_in_at_least_one_phi} to be true and the RHS to be false, then:
\begin{align*}
    & \lor_{i\in\mathcal{I}}\left( \dot{h}^F_i(x(t),u^*(x(t),d^*(x(t)))) \leq 0\right) = \mathrm{FALSE} & \implies \\
    & \wedge_{i\in\mathcal{I}}\left( \dot{h}^F_i(x(t),u^*(x(t),d^*(x(t)))) > 0\right) = \mathrm{TRUE} & \implies \\
    & \sum_{i\in\mathcal{I}}\dot{h}^F_i(x(t),u^*(x(t),d^*(x(t)))) > 0,
\end{align*}
which is a contradiction.

As a result, $d^*(x(t))$ attempting to ensure equation~\eqref{eqn::multi_CBF_inequality_forall_time} is equivalent to saying $d^*(x(t))$ attempts to ensure:
\begin{multline}
    \lor_{i\in\mathcal{I}}\left( \dot{h}^F_i(x(t),u^*(x(t),d^*(x(t)))) \leq 0\right) \lor \label{eqn::d_specification_CBF} \\
     \lor_{j\in\mathcal{J}}\bigg(\dot{h}^G_j(x(t),u,d^*(x(t))) <  -\alpha_j(h^G_j(x(t),d^*(x(t))),  \\
    ~\forall~u\in\mathcal{U}\bigg),  \forall~t\geq 0.
\end{multline}
Coupled with the initial conditions~\eqref{eqn::initially_not_F} and \eqref{eqn::initially_G}, $d^*(x(t))$ attempting to ensure equation~\eqref{eqn::d_specification_CBF} is equivalent to saying $d^*(x(t))$ attempts to ensure system satisfaction of equation~\eqref{eqn::d_specification}, which is the desired result.
\end{proof}

\section{Test Synthesis}
\label{sec::corollaries}
This section provides some additions to the main result that make it extensible to the problem at hand.  Specifically, we formulate a linear program to extend the results of Theorem~\ref{algorithmic_test_generation} to generate test cases wherein we have no prior knowledge of the controller on-board the system.  Likewise, we have a corollary that permits a predictive form of equation~\eqref{differential_game} such as the one used to generate the tests in Section~\ref{sec::simulations_and_experiments}.

To start, we want to use the results of Theorem~\ref{algorithmic_test_generation} to see if the provided autonomous system satisfies the associated TL specification.  However, as we do not know the controller onboard the system, we do not have any CBFs with which to define the minimax game in Theorem~\ref{algorithmic_test_generation}.  That being said, Lemma~\ref{equivalence_logic_safety} in the Appendix provides us a method by which to determine these CBFs from the system demonstration data, $\mathbb{D}$.  First, we define an estimated CBF (e-CBF) to be a convex combination of component functions in $\mathcal{B}$, where $p_j$ below denote the weights for said combination:
\begin{equation}
    \label{optimal_combination} \tag{e-CBF}
    h^*(x) = \sum_{j=1}^{|\mathcal{B}|} p_jh_j(x), \quad \forall~h_j \in \mathcal{B}.
\end{equation}

By default, Lemma~\ref{equivalence_logic_safety} indicates that specification satisfaction requires the associated CBF to be positive.  As a result, we will choose a cost function that is minimized when \eqref{optimal_combination} is most positive over all demonstrations:
\begin{equation}
    \label{eqn::cost}
    J(\mathcal{B},x,p) \triangleq -\sum_{i=1}^{|\mathbb{D}|} \sum_{k=0}^{T_i} \sum_{j=1}^{|\mathcal{B}|} p_j h_j(x^i_k). \tag{Cost}
\end{equation}
Likewise, Assumption~\ref{Labeling} dictates that demonstrations end upon satisfaction of the control objective.  Therefore, for the estimated CBF to correspond to $\F$uture type constraints, \eqref{optimal_combination} should be positive at the end of each demonstration.  Similarly, for $\G$lobal type constraints, \eqref{optimal_combination} should be positive over the length of all demonstrations.  This results in the following Corollary:

\begin{figure*}[htbp]
    \centering
    \includegraphics[width = 0.99\textwidth]{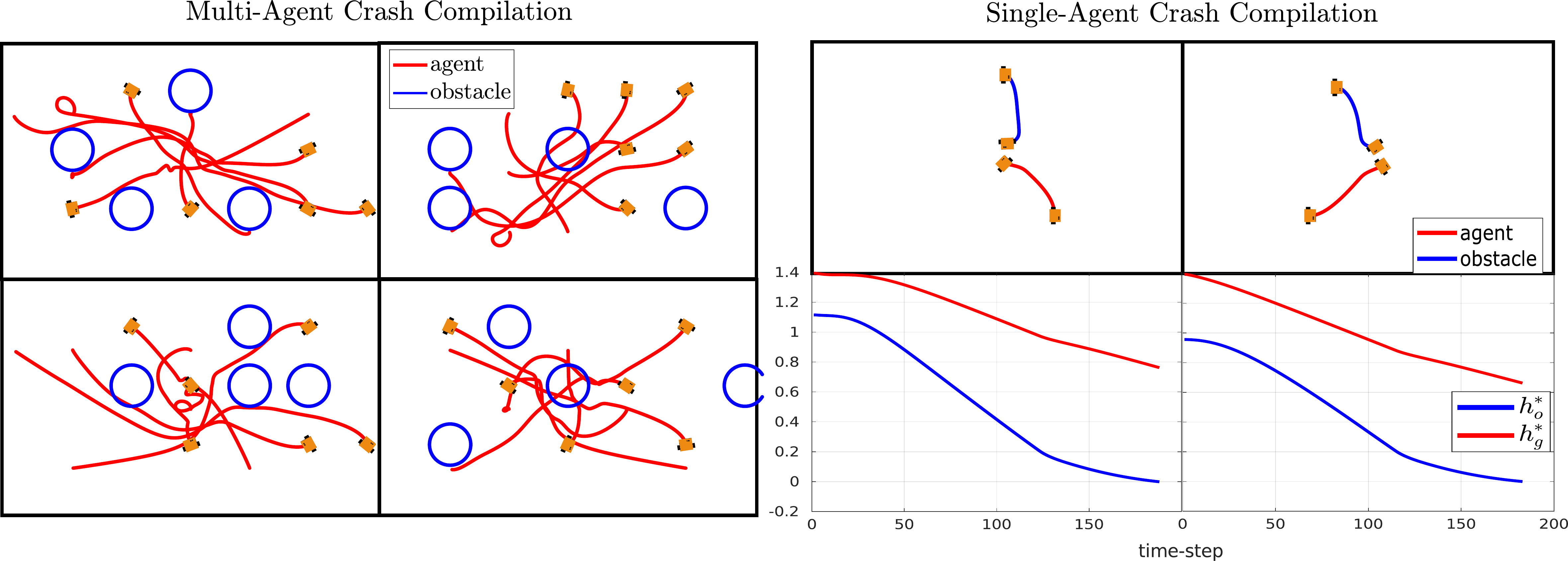}
    \caption{Shown above are simulated examples of tests generated by the test generation framework detailed in the paper.  In all cases shown, each agent's goal is to cross the map to a location directly opposite to its starting location.  Here, the origin is the center of each rectangular region. In all figures, red lines denote agent trajectories and blue lines denote obstacle trajectories/regions if stationary.  (Left) Two, stacked figures showing successful robot navigation with multiple, stationary obstacles - this is our simulated demonstration data for the multi-agent case.  (Center Left) Two, stacked figures which show stationary obstacle placement based on the test framework, note that, in the simulations shown, multiple crashes occur.  (Center Right and Right) Two simulations of a single agent in a moving obstacle case.  In both cases, the obstacle trajectory is updated to match the obstacle location generated by minimax game~\eqref{experiment_game}.  Note that in both cases, the simulation terminates when obstacle safety has been violated ($h^*_o \leq 0$).}
    \label{fig::simulation_crashes}
\end{figure*}
\vspace{.2cm}
\begin{corollary}
    \label{composite_CBF_Lemma}
\textit{    For a given data-set, $\mathbb{D}$, and a candidate set of functions, $\mathcal{B}$, the solution, $p^*$, to the following linear program:
    \begin{align}
        p^* = &\,\,\,\argmin\limits_{p\in\mathbb{R}^{|\mathcal{B}|}} & & J(\mathcal{B},x,p), && \label{CBF-LP} \tag{CBF-LP} \\ 
        & \subjectto & & \eqref{future-constraint}~\mathrm{or}~\eqref{global-constraint}, \nonumber  && \\
        & & & p_j \geq 0, \quad \forall~j=1,2,\dots,|\mathcal{B}|,  && \nonumber\\
        & & & \sum_{j=1}^{|\mathcal{B}|} p_j = 1, && \nonumber
    \end{align}
    \begin{subequations}
    \begin{align}
        \label{future-constraint}
        &\hspace{-0.3in}\sum_{j=1}^{|\mathcal{B}|} p_j h_j(x^i_{T_i}) & & \hspace{-0.3in}\geq 0,~\forall~i=1,2,\dots,r, \\
        \label{global-constraint}
        &\hspace{-0.3in}\sum_{j=1}^{|\mathcal{B}|} p_j h_j(x^i_k) & & \hspace{-0.3in} \geq 0,~ \forall~i=1,\dots,r,~k=0,1,\dots,T_i,
    \end{align}
    \end{subequations}
    determines an estimated CBF, \eqref{optimal_combination}, for specifications of type $\F\phi$ (constraint~\eqref{future-constraint}) or type $\G\phi$ (constraint~\eqref{global-constraint}).
    Furthermore, for $\F\phi$:
    \begin{equation}
        \mathcal{C}_{h^*} \cap \llbracket \phi \rrbracket \supseteq \{x^i_{T_i}\}~\forall~i=1,2,\dots,r,
    \end{equation}
    and for $\G\phi$:
    \begin{equation}
        \mathcal{C}_{h^*} \cap \llbracket \phi \rrbracket \supseteq \{x^i_k\}~\forall~k = 0,1,\dots,T_i~\mathrm{and}~i=1,\dots,r.
    \end{equation}
    }
\end{corollary}
\vspace{.2cm}
\begin{proof} 
    Assumption~\ref{Labeling} requires that $\phi =$ TRUE at each $T_i$ for $\mathbb{D}_i$.  If a solution to equation~\eqref{CBF-LP} exists with constraint \eqref{future-constraint}, then we have:
    \begin{align*}
        x^i_{T_i} & \in \llbracket \phi \rrbracket, \\
        h^*(x^i_{T_i}) & = \sum_{j=1}^{|\mathcal{B}|} p_j^*h_j(x^i_{T_i}) \geq 0,
    \end{align*}
    for any solution, $p^*$.
    As a result,
    \begin{align*}
        \mathcal{C}_{h^*} \cap \llbracket \phi \rrbracket \supseteq \{x^i_{T_i}\} \quad \forall~i=1,2,\dots,r,
    \end{align*}
    which only implies set equivalence up to the provided data.  As a result, Lemma~\ref{equivalence_logic_safety} only applies over the provided data-set where the equivalence holds.  To show the same for $\G$lobal type specifications, replace constraint \eqref{future-constraint} with \eqref{global-constraint} and the proof follows similarly. 
\end{proof}
\vspace{0.2 cm}

As the CBFs generated via Corollary~\ref{composite_CBF_Lemma} are not exact, the results of Theorem~\ref{algorithmic_test_generation} cannot be guaranteed.  However, they are very useful in generating tests as will be shown in Section~\ref{sec::simulations_and_experiments}.  

Secondly, the minimax game in Theorem~\ref{algorithmic_test_generation} may be non-convex and/or calculation of the solution may be computationally difficult.  However, as the minimax game depends only on the current state, we can calculate the optimal test parameters for some subset of points and define the actual test to be an interpolation of the parameters defined at these points.  That being said, this yields sub-optimal tests.  
Finally, the minimax problem~\eqref{differential_game} need not have that specific cost function for it to determine optimal test parameters.  It suffices if the chosen cost function for the inner maximization problem is maximized when the estimated CBFs $h^F_i(x) \geq 0$.  This permits predictive games of the form used in the simulations in Section~\ref{sec::simulations_and_experiments}.

\section{Case Study}
\label{sec::simulations_and_experiments}
In this section, we detail simulations of test scenarios devised by our framework  applied to the Georgia Tech Robotarium \cite{Robotarium}. The set up is defined next.

\newsec{System Specification} $\mathbf{F}g_i \wedge_j \mathbf{G}\neg a_j$.  The system will always ensure that agent $i$ ends up at goal location $i$ while avoiding all the other agents $j$, provided that $\llbracket g_i \rrbracket \not \subseteq \cup_j \llbracket a_j \rrbracket$.
For this specification, we have the following information:
\begin{itemize}
    \item $\mathbb{D}$: A set of twenty demonstrations ($r=20$) derived from simulations wherein a single agent successfully navigates to a predetermined goal while avoiding another obstacle.
    \item $\mathcal{B}_F$: A set of norm-based barrier functions of the form $h(x) = -\|x - c\| + r$ wherein $h(x) \geq 0 \equiv \|x-c\| \leq r$.
    \item $\mathcal{B}_G$: A set of norm-based barrier functions of the form $h(x) = \|x - c\| - r$ wherein $h(x) \geq 0 \equiv \|x-c\| \geq r$.
\end{itemize}
Equation~\eqref{CBF-LP} identified:
\begin{align*}
    h^*_g(x) &=  - \|x-g\| + 0.02, \\
    h^*_o(x,d) &= \|x-d\| - 0.175,
\end{align*}
as the estimated CBF's for $\F g_i$ and $\G \neg a_j$ respectively.  Here, $d$ is the desired location of the obstacle agent, and the testing parameter we control.  We estimated that the robots in the Robotarium can be sufficiently modeled with single-integrator systems and developed the following game from the derived barrier functions:
\begin{align}
    d^* = &  \, \, \, \, \argmin\limits_{d \in \mathbb{R}^2} &  \max\limits_{u \in \mathcal{U}} & \, \,  \sum_{i=1}^{N}h^*_g(x_i), 
    \label{experiment_game} \tag{Simulation Game} \\
    & \subjectto & & \, \, \dot{h}^*_o(x_{i-1},u_i,d) \geq -\beta h^*_o(x_{i-1},d), \nonumber \\
    & & & \, \, x_i = x_{i-1} + u_i\Delta t, \nonumber\\
    & & & \,\, \dots\forall~i=1,2,\dots,N, \nonumber \\
    & & & \, \, \|d-x_0\| \geq r_o. \nonumber 
\end{align}
In equation~\eqref{experiment_game} above, $N=2$, $\beta = 100$, and $r_o = 0.175$. For large values of $\beta$, there is less of an implied assumption about system behavior as it decays to the boundary of the estimated safe region.  As a result, large $\beta$ values permit equation~\eqref{experiment_game} to account for a wider range of system behavior when solving for $d^*$.  In addition,  $r_o$ constrains against trivial solutions wherein $d=x_0$, which makes the inner maximization problem infeasible.

To quantify how "hard" a test/demonstration is, we define:
\begin{itemize}
    \item $H^i_g \triangleq \frac{1}{T_i+1} \sum_{k=0}^{T_i} \left| \hat{h}_g(x_k)\right|$ to be the average time the system spent outside the goal.  Here, $\hat{h}_g$ denotes a normalized version of our estimated CBF, $h^*_g$, such that $-1 \leq \hat{h}_g(x_k) \leq 0$, $\forall$ $k = 0,1,\dots T_i$, and $T_i$ is the max time for our Demonstrations/tests as defined in Definition~\ref{demonstrations}.
    \item $H^i_o \triangleq 1 - \frac{1}{T_i+1} \sum_{k=0}^{T_i} \hat{h}_o(x_k)$ to be the average time spent collision free.  Here, $\hat{h}_o$ denotes a normalized version of our estimated CBF, $h^*_o$ such that $0 \leq \hat{h}_o(x_k) \leq 1$, $\forall$ $k = 0,1,\dots,T_i$.
\end{itemize}
To note, tests drive $H^i_o \to 1$ in an effort to drive $H^i_g \to 1$ which denotes system safety failure and inability to reach the goal, respectively.

\begin{figure}[t]
    \centering
    \includegraphics[width = 0.45\textwidth]{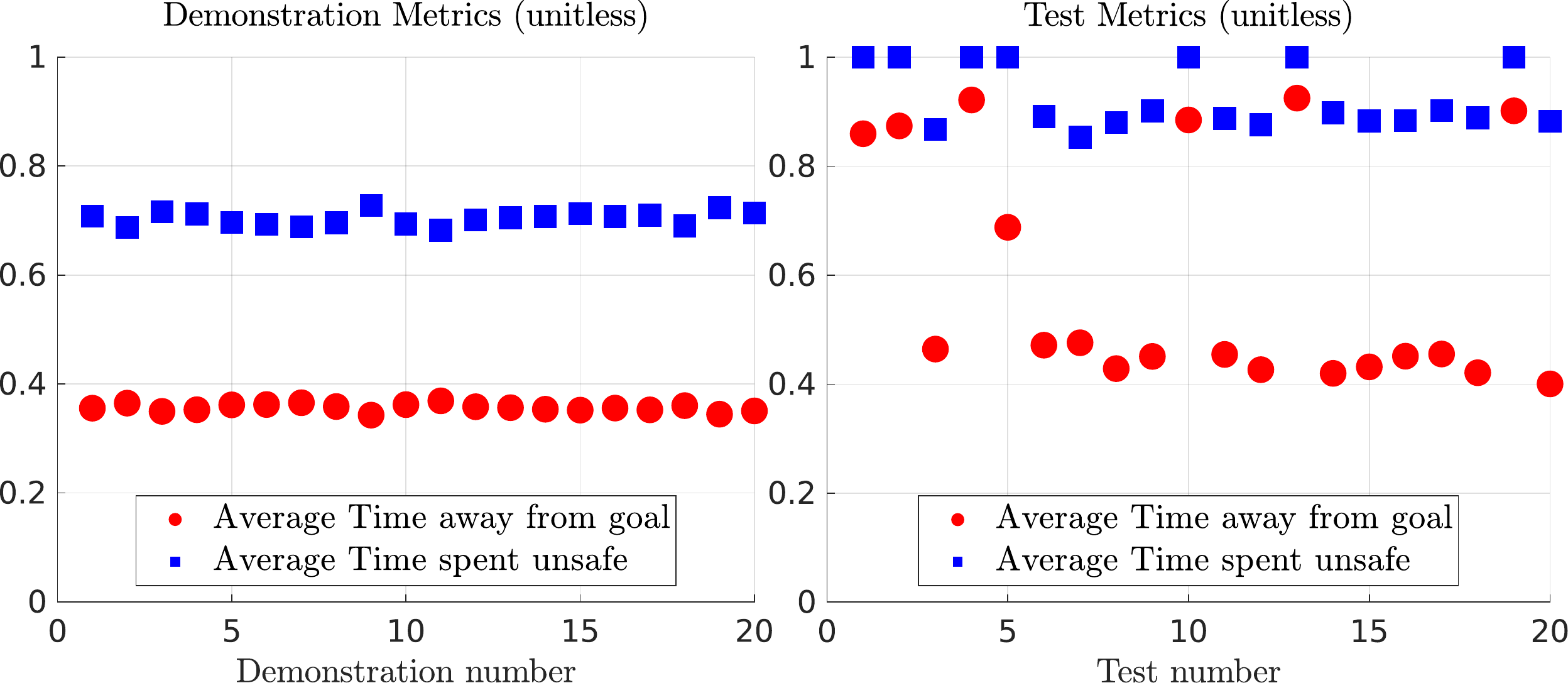}
    \caption{The figures compares the specification satisfaction performance of the provided demonstration data (left) with the data generated during simulated tests (right) based on the metric defined in Section V.  Note that for all test cases, the agent took longer to reach its goal, and in $7/20$ cases, the test caused the robots to crash $H^i_o = 1$.}
    \label{fig::Metrized Data}
\end{figure}

Figures~\ref{fig::simulation_crashes} and \ref{fig::Metrized Data} show the results of simulations based on the obstacle locations outputted by minimax game~\eqref{experiment_game}.  For the multi-agent case, examples of the provided demonstration data are shown in the two, stacked figures to the far left.  Under normal operating parameters, the agent successfully avoid the obstacles while moving to their repsective goals (none of the red lines interset the blue circles).  However, when the stationary obstacle locations are updated based on solutions to \eqref{experiment_game}, multiple crashes occur as shown in the two, stacked figures just left of center.  

Likewise, for the single-agent simulations shown, we inputted the desired obstacle location, $d$, as the goal location for a secondary agent.  This agent acted as a moving obstacle, and for $2/20$ tests simulated, the trajectories for both agents are shown in the four figures on the right of Figure~\ref{fig::simulation_crashes}.  In both of these cases, notice how the estimated CBF, $h^*_o$, decays to $0$ upon termination.  Effectively, in both of these cases, the test framework chose a sequence of obstacle locations, $d^*(x(t))$, that forced the system to satisfy $\G a_j$, at least with respect to the estimated CBF, $h^*_o$.  

Data for all $20$ single-agent simulations are compared against the provided data-set, $\mathbb{D}$, in Figure~\ref{fig::Metrized Data}.  Under normal operation, the demonstration data is relatively consistent \textit{i.e.} $H^i_g$ hovers just below $0.4$ and $H^i_o$ hovers just around $0.7$ for all demonstrations, $i=1,2,\dots,20$.  However, for all test simulations, $H^i_g > 0.4$ and $H^i_o > 0.8$ further corroborating that the test parameter sequence generates difficult tests, and in $7/20$ cases wherein $H^i_o=1$, also forced the system to satisfy specification~\eqref{eqn::d_specification}.  An example of an experimental demonstration of the test framework can be seen in an accompanying video (linked here: \cite{video}).  The setup here mimics the same single-agent case shown in the examples in Figure~\ref{fig::simulation_crashes}.

\section{Conclusion}
In this paper, we attempt to solve the problem of test and evaluation for verification and validation of autonomous systems, wherein the specific controllers are unknown.  The goal in doing so, is to provide a mathematical framework designed to root out system inefficiencies in an effort to ensure confidence in those systems that pass the procedure.  The method detailed involves estimation of approximate control barrier functions to frame a minimax game that is guaranteed to choose test parameters to frustrate system satisfaction of a provided temporal logic specification.  In the future, we aim to extend this work to richer specification classes and formalize an iterative testing procedure based on our framework.

\bibliographystyle{ieeetr}
\bibliography{collected_works}

\end{document}